\documentclass[aip,pop,reprint,groupedaddress]{revtex4-1}
\usepackage{amsmath}
\usepackage{amssymb}
\usepackage{graphicx}
\usepackage{bm}

\begin{document}

\title{Control of focusing forces and emittances in plasma-based
accelerators using near-hollow plasma channels}

\author{C. B. Schroeder}
\author{E. Esarey}
\author{C. Benedetti}
\author{W. P. Leemans} 

\affiliation{Lawrence Berkeley National Laboratory, Berkeley,
California 94720 USA}

\date{\today}

\begin{abstract}
A near-hollow plasma channel, where the plasma density in the channel
is much less than the plasma density in the walls, is proposed to
provide independent control over the focusing and accelerating forces
in a plasma accelerator.  In this geometry the low density in the
channel contributes to the focusing forces, while the accelerating
fields are determined by the high density in the channel walls.  The
channel also provides guiding for intense laser pulses used for
wakefield excitation.  Both electron and positron beams can be
accelerated in a nearly symmetric fashion.  Near-hollow plasma
channels can effectively mitigate emittance growth due to Coulomb
scattering for high-energy physics applications.
\end{abstract}

\maketitle

\section{Introduction}

Plasma-based accelerators are of interest because of their ability to
sustain large acceleration gradients, enabling compact accelerating
structures.  The electric field of the electron plasma wave is on the
order of $E_0 = cm_e\omega_p/e$, or $E_0 [{\rm V/m}] \simeq 96
\sqrt{n_0 [{\rm cm}^{-3}]}$, where $\omega_p = k_p c = (4\pi n_0
e^2/m_e)^{1/2}$ is the electron plasma frequency, $n_0$ is the ambient
electron number density, $m_e$ and $e$ are the electron rest mass and
charge, respectively, and $c$ is the speed of light in vacuum.  This
field can be several orders of magnitude greater than conventional
accelerators, which are limited by material breakdown.  Electron
plasma waves with relativistic phase velocities may be excited by the
ponderomotive force of an intense laser \cite{Esarey09} or the
space-charge force of a charged particle beam.
\cite{PisinChen85,Rosenzweig88} High-quality 1~GeV electron beams have
been produced using 40~TW laser pulses in cm-scale plasmas.
\cite{Leemans06b} Beam-driven plasma waves have also been used to
double the energy of a fraction of electrons on the beam tail by the
plasma wave excited by the beam head.  \cite{Blumenfeld07} These
experimental successes have resulted in further interest in the
development of plasma-based acceleration as a basis for future linear
colliders.  \cite{Seryi09,*Adil13,Schroeder10b,Schroeder12c}

The focusing forces acting on beams originate from the transverse
wakefields in the plasma.  For beam or laser drivers in the blow-out
regime the focusing forces are determined by the background ion
density.  \cite{Esarey09} For laser-driven plasma waves in the
quasi-linear regime, the focusing forces can be controlled by
controlling the transverse wakefields using shaped transverse laser
intensity profiles.\cite{Cormier-Michel11} Matching the beam to the
focusing force is required to prevent emittance growth.  Independent
control of the transverse focusing and longitudinal accelerating
forces is desired for control of the beam radius, enabling matched
propagation.\cite{Assmann98}

For high-energy physics applications, beams require ultra-low
emittance to achieve the required luminosity.  For fixed beam size,
matching the beam in a plasma accelerator requires adjusting the
focusing force of the wakefield as the beam accelerates such that
$k_\beta = \epsilon_n/(\gamma\sigma_x^2)$, where $k_\beta$ is the
betatron wavenumber of the focusing force, $\epsilon_n$ is the
normalized transverse emittance, $\sigma_x$ the beam transverse size,
and $\gamma m_e c^2$ is the beam particle energy.  For a beam density
greater than the plasma density $n_b \gg n_0$, the beam will blowout
the surrounding electrons such that $k_\beta = k_p/\sqrt{2\gamma}$,
and for sufficiently high beam density in a uniform plasma $n_b/n_0
\gg M_i/m_e$, where $M_i$ is the ion mass, the background plasma ions
will move on the plasma period time-scale, leading to emittance
growth.  \cite{Rosenzweig05} Maintaining $n_b \lesssim n_0$ typically
requires weak focusing $k_\beta = \epsilon_n/(\gamma\sigma_x^2) \ll
k_p/\sqrt{\gamma}$.  This weak focusing can lead to emittance growth
via Coulomb scattering with background ions in the plasma.

Hollow plasma channels, with zero density out to the channel radius
and constant density for larger radii, have been studied
\cite{Chiou95,Chiou98,Schroeder99a} due to the beneficial properties
of the accelerating structure.  In a hollow plasma channel, the transverse
profile of the driver is largely decoupled from the transverse profile
of the accelerating mode.  Therefore, for a relativistic driver, the
accelerating gradient is transversely uniform and the focusing fields
are linear.  In addition, the accelerating mode of the hollow plasma channel
is primarily electromagnetic, unlike the electrostatic fields excited
in a homogeneous plasma.  Methods for hollow plasma channel creation
are actively being explored.  \cite{Kimura11}

In this work, we propose to use a partially-filled (near-hollow)
plasma channel, with plasma density in the channel much less than the
plasma density in the wall, to provide independent control of
the focusing forces.  In this geometry, the plasma density in the
channel contributes to the focusing force, and the accelerating force
is determined by the plasma density in the wall.  For a
sufficiently relativistic driver, the focusing wakefield is
transversely linear and axially uniform.  Hence any projected
transverse emittance growth due to beam head-to-tail mismatch is
eliminated.  \cite{Mehrling12} It is also shown that the accelerating
and focusing forces in this geometry can mitigate emittance growth
induced by Coulomb collisions with background ions, preserving
ultra-low emittance for high-energy physics applications.  Ion motion
in the channel is also negligible for resonant beams.  A near-hollow
channel, as described below, allows matched beam propagation with a
constant beam density without significant emittance growth via
scattering.  Both electrons and positrons beams can be accelerated in
a nearly symmetrical fashion.

\section{Wakefields in near-hollow plasma channels}
 \label{sec:wake}

Consider a plasma channel with an initial electron plasma density of
the form $n (r) = n_c$ for $r < r_c$ and $n (r) = n_w$ for $r\geq
r_c$, where $n_w$ is the density in the wall, $n_c$ is the density in
the channel ($n_c\ll n_w$), and $r_c$ is the channel radius.  We will
consider $k_w r_c \sim 1$ where $k_w^2 = 4 \pi n_w e^2/m_ec^2$ is the
plasma wavenumber corresponding to the wall density.  To provide weak
focusing of an electron beam we will consider $k_c^2 \ll k_w^2$, where
$k_c^2 = 4 \pi n_c e^2/m_ec^2$ is the plasma wavenumber corresponding
to the channel density.

Excitation of large amplitude plasma waves requires high-intensity
lasers, $a_0\sim 1$, where $a_0^2 \simeq 7.32 \times 10^{-19}
\lambda_0^2[\mu\textrm{m}] I_0 [\textrm{W/cm}^2]$ with $\lambda_0$ the
laser wavelength and $I_0$ the peak laser intensity.  Note that such a
plasma channel can effectively guide a laser pulse, \cite{Chiou95} as
the channel depth (not the on-axis density) provides for laser
guiding.  \cite{Esarey09} The laser spot size $w_0$ for quasi-matched
propagation can be computed following Ref.~\onlinecite{Benedetti12},
and $w_0 \sim r_c > k_w^{-1}$ is typically required for guiding.  
For example, guiding for a transversely Gaussian laser pulse is
provided in the low intensity, low power limit for a laser spot size
$w_0 = r_c/[\ln (k_w r_c)]^{1/2}$.
In this
regime $a_0^2/(1+a_0^2/2)^{1/2} > k_c^2 w_0^2 /2$, and the transverse
ponderomotive force of the laser will expel the channel electrons,
leaving an ion column.  \cite{Esarey09} This ion column can provide
linear, phase-independent focusing for an ultra-relativistic witness
bunch.  In the following we will consider wakefields excited by a
laser driver, although a particle beam driver can also excite
wakefields with similar properties in a near-hollow plasma channel.

The wake excited in such a channel will consists of an electromagnetic
wake owing to surface currents driven in the channel walls and a wake
owing to the background ions in the channel.  In the limit $k_c^2 \ll
k_w^2$, the accelerating field in the channel is dominated by the
currents in the wall and has the form \cite{Chiou95}
\begin{equation}
E_z \simeq -  E_w \Omega \int_\infty^{\zeta} k d\zeta'
\cos [k (\zeta-\zeta') ] a^2( r= r_c, \zeta')/4
,
\label{eq:ez}
\end{equation}
where $a (r, \zeta ) = eA/m_ec^2$ is the normalized transverse vector
potential profile of the laser, $\zeta = z-\beta_p c t$, $\beta_p c$
is the driver velocity and $\gamma_p = (1-\beta_p^2)^{-1/2} \gg 1$.
For a laser driver, $\gamma_p \sim k_0/k_w = 2\pi/(k_w\lambda_0)$.
The excited mode wavenumber\cite{Chiou95,Schroeder99a} is $k = k_w
\Omega$ with
\begin{equation}
\Omega = \left[1+ \frac{k_wr_c K_0(k_w r_c)}{2K_1(k_w r_c)} \right]^{-1/2}
,
\end{equation}
and, for typical parameters, $\Omega \sim 1$.
The focusing field excited in the channel
is given by
\begin{multline}
E_r - \beta B_\theta \simeq  E_c \frac{k_c r}{2} 
- E_w \frac{k_w r}{4}    \left( \gamma^{-2} + \gamma_p^{-2} \right)
\\ \times
\Omega^2
\int_\infty^{\zeta} k d\zeta' \sin [k (\zeta-\zeta') ] a^2( r= r_c, \zeta')/4
,
\label{eq:fr}
\end{multline}
where $\gamma^2 = 1/(1-\beta^2) \gg 1$ and $c\beta$ is the witness
beam velocity.  Here $E_w = m_e c^2 k_w/e$ and $E_c = m_e c^2 k_c/e$.
The first term on the right-hand side of Eq.~\eqref{eq:fr} is due to
the ion column and the second term is due to the currents driven in
the channel walls.  The focusing force is linear, to order
$\mathcal{O}(\gamma_p^{-2})$, with respect to the radial position $E_r
- \beta B_\theta \propto r$ and hence the rms normalized transverse
(slice) emittance is conserved to that order.  In the regime $n_c/n_w
> a_0^2(r_c)/(8\gamma_p^2)$ and $\gamma^{2} \gg \gamma_p^2$, the
focusing from the channel ion density dominates $E_r - \beta B_\theta
\simeq E_c k_c r / 2 $ and $k_\beta = k_c/\sqrt{2\gamma}$.  In this
case the focusing force is uniform (in phase) over the entire bunch,
eliminating any betatron mismatch between the head and tail of the
beam.\cite{Mehrling12} Matched propagation is achieved for 
\begin{equation}
\frac{n_c}{n_w} = \frac{2 (k_w \epsilon_n)^2}{\gamma (k_w \sigma_x)^4}
.
\end{equation}  
For the case of an effectively hollow plasma channel, the focusing forces can
be controlled by using an external (permanent magnet) focusing system.
Acceleration of positron beams would operate in this regime.

\begin{figure}
\begin{center}
\includegraphics[scale=1]{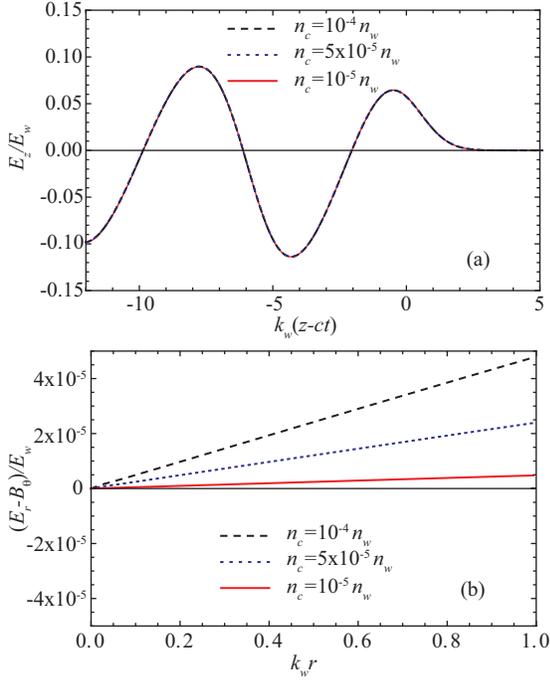}
   \end{center}
\caption{
\label{fig:wake}
Plasma wakefield excited by a quasi-matched, resonant Gaussian laser
pulse with $a_0=1$ and $k_w w_0 = 2.3$ in a near-hollow plasma channel
with $k_w r_c = 1.5$ and $n_c/n_w = 10^{-5}$, $5\times 10^{-5}$ and
$10^{-4}$.  (a) Accelerating wakefield $E_z/E_w$ versus $k_w (z-ct)$.
(b) Focusing wakefield $(E_r-B_\theta )/E_w$ in the channel (at the
peak $E_z$) versus $k_w r$.}
\end{figure}

Figure~\ref{fig:wake} shows the normalized longitudinal $E_z/E_w$ and
transverse $(E_r-B_\theta)/E_w$ wakefields excited by a
quasi-matched,\cite{Benedetti12} resonant Gaussian laser with $a_0=1$,
$k_0/k_w = 100$, and $k_w w_0 = 2.3$, in a channel with $k_w r_c=1.5$
and $n_c/n_w = 10^{-5}$, $5\times 10^{-5}$ and $10^{-4}$.  The
wakefields shown in Fig.~\ref{fig:wake} were computed using the
particle-in-cell code {\tt INF\&RNO}.\cite{Benedetti10,*Benedetti12b}
Figure~\ref{fig:wake}(a) shows the accelerating field (determined by
the wall density) is independent of the channel density for $n_c \ll
n_w$.  In addition, the accelerating field is uniform with respect to
radial position inside the channel.  Figure~\ref{fig:wake}(b) shows
independent control over the focusing field on a witness bunch (for
fixed accelerating field) by varying the channel density.  These
transverse wakefields have excellent properties for beam emittance
preservation, namely linear $\propto r$ and axially constant
throughout a witness beam.

\section{Emittance growth by scattering}
\label{sec:scatt}

It is well-known that emittance growth can occur by elastic scattering
with the plasma ions.  \cite{Montague85,Kirby07,Schroeder10b} Coulomb
collisions results in a change of the rms divergence of the beam
particle.  \cite{Nicholson92} For the case of a near-hollow plasma channel,
\begin{multline}
\frac{d\langle \Delta\theta_x^2 \rangle}{dz} = \frac{4\pi r_e^2}{
\gamma^{2}} \int^{r_{\rm max}}_{ r_{\rm min}} \frac{dr}{r} {Z^2 n_{i}} 
\\
= 
\frac{r_e}{ \gamma^{2}}\left[  
Z_c k_c^2  \ln \left( \frac{r_c}{ r_{\rm min}}\right) 
+ Z_w k_w^2 \ln \left( 1 +\frac{ \lambda_D}{r_c}\right)
\right]
,
\label{eq:scat}
\end{multline}
where $\Delta \theta_x = \Delta p_x/p_z$ is the ratio of perturbed
transverse particle momentum to longitudinal particle momentum (the
brackets indicate an rms average over many scattering events),
$Z_{c,w}$ is the charge state of the ions, and $r_e = e^2/m_ec^2$.  In
the quasi-linear wakefield regime $a^2 (r=r_c)<1$, background
electrons are present and provide screening in the channel walls, such
that the maximum impact parameter is $r_{\rm max} \sim r_c +
\lambda_{\rm D}$ where $\lambda_{\rm D} = (k_{\rm B}T/4\pi n_w
e^2)^{1/2}$ is the Debye length.  For typical laser-ionized plasmas,
the electron plasma temperature $k_{\rm B}T$ is on the order of eV.
For $k_w r_c \sim 1$, $ \lambda_D/r_c \sim (k_{\rm B}T /m_ec^2)^{1/2}
= \beta_{\rm th} \ll 1$.  The minimum impact parameter is the
effective nuclear radius, $r_{\rm min} = 1.4 A^{1/3}$~fm, with $A$ the
mass number.\cite{Jackson75} For typical parameters, $\ln ( {r_c}/{
r_{\rm min}} )\sim 10$ and $\ln ( 1 + \lambda_D/r_c ) \sim \beta_{\rm
th}\sim 10^{-3} $.
For a relativistic particle undergoing linear focusing
($F_x/\gamma m_ec^2 = -k_\beta^2 x$) with acceleration and an
approximately matched beam, the resulting rms normalized
emittance $\epsilon_n = \gamma \epsilon_x $ growth is
\begin{equation}
  \frac{d{\epsilon_n}}{dz} = 
\frac{\gamma}{2k_\beta} \frac{d\langle \theta_x^2 \rangle}{dz}
.
\end{equation}
Strong focusing (large $k_\beta$) suppresses the emittance growth from
scattering.

\subsection{Constant focusing force in homogeneous plasma}

To compare to the case of a near-hollow plasma channel, we first review the
case of scattering in a homogeneous plasma.
\cite{Montague85,Kirby07,Schroeder10b} Consider a constant focusing
force, with $k_\beta = \kappa/\gamma^{1/2}$ and $\kappa =
\textrm{constant}$.  For the quasi-linear laser-driven wake regime,
$\kappa^2 = (\phi_0/r_\perp^2) \sin \psi$, where $\psi$ is phase
location of the beam in the plasma wave, $r_\perp$ is the transverse
scale length of the wakefield, and $\phi_0$ is the wakefield
amplitude.  The accelerating field is $d{\gamma}/dz = -k_p E_z/E_0$.
For the quasi-linear laser-driven wake regime, $d{\gamma}/dz = k_p
\phi_0 \cos \psi$.  In the highly-nonlinear bubble
regime,\cite{Kirby07} $\kappa = k_p/\sqrt{2}$ and $E_z/E_0 = k_p r_B
/2$, where $r_B$ is the bubble radius (approximately the nonlinear
plasma wavelength).

For constant wakefield focusing, the emittance growth is
\begin{equation}
   \frac{d\epsilon_n}{d\gamma} = 
   \frac{ k_p r_e Z 
}{2 \kappa (E_z/E_0)\gamma^{1/2}} 
\ln \left( \frac{r_{\rm max}}{ r_{\rm min}} \right) 
, 
\end{equation}
which may be solved to yield
\begin{equation}
   \Delta \epsilon_{n}  = 
   \left[ \frac{ k_p }{\kappa (E_z/E_0)} \right] r_e Z\ln \left(  
\frac{r_{\rm max}}{ r_{\rm min}}\right) 
\left( \gamma_f^{1/2} - \gamma_i^{1/2} \right) 
,
\label{eq:emit-homo}
\end{equation}
where $\Delta \epsilon_n = \epsilon_{nf} - \epsilon_{ni}$ with
$\epsilon_{nf}$ and $\epsilon_{ni}$ the final and initial normalized
emittances, respectively, and $\gamma_f$ and $\gamma_i$ are the final
and initial beam energies, respectively.  Note that $r_{\rm max} =
\lambda_{\rm D}$ in the quasi-linear regime and $r_{\rm max} = r_B$ in
the nonlinear bubble regime.  
Typically $k_p r_\perp \sim 1$ and $k_p r_{B} \sim 1$, such that
$k_p/[\kappa (E_z/E_0)] \sim 1$ and there is a weak dependence on the
plasma density.  Scattering is suppressed by the strong focusing
provided by the plasma wave.  As the beam accelerates, the beam radius
decreases and the peak beam density increases $n_b \propto
\gamma^{1/2}$.  For sufficiently high beam density $n_b/n_0 \gtrsim
M_i/m_e$, ion motion will occur.  \cite{Rosenzweig05}

\subsection{Constant beam density in homogeneous plasma}

If fixed beam density is desired, i.e., $\sigma_x =
\textrm{constant}$, one may consider varying the focusing force with
energy such that the beam remains matched $k_\beta =
{\epsilon_n}/({\gamma\sigma_x^2})$ or $ \kappa \propto
{1}/{\gamma^{1/2}} $.  In this case, the emittance growth is
\begin{equation}
   \frac{d\epsilon_n}{d\gamma} = 
   \frac{ \sigma_x^2 k_p r_e Z 
}{2\epsilon_n (E_z/E_0)} 
\ln \left( \frac{\lambda_{\rm D}}{ r_{\rm min}} \right) 
, 
\end{equation}
which may be solved to yield
\begin{equation}
   \epsilon_{nf}^2 - \epsilon_{ni}^2 = 
   \frac{ k_p r_e Z \sigma_x^2 
}{  (E_z/E_0)} \ln \left(  \frac{\lambda_{\rm D}}{ r_{\rm min}} \right) 
\left( \gamma_f - \gamma_i \right) 
.
\end{equation}
With $\gamma_f \gg\gamma_i$
and $\epsilon_{nf} \gg\epsilon_{ni}$,
\begin{equation}
   \epsilon_{nf} \simeq
   \left[ \frac{ k_p r_e Z \sigma_x^2 \gamma_f 
}{  (E_z/E_0)} \ln \left(  \frac{\lambda_{\rm D}}{ r_{\rm min}} \right) 
\right]^{1/2}  
.
\label{eq:emit-const-n_b}
\end{equation}
This emittance growth can be prohibitively large for high-energy
physics applications.  Note that a similar expression was obtained in
Ref.~\onlinecite{Lebedev13}.

\subsection{Constant beam density in near-hollow plasma channel}

Consider the case of scattering in a near-hollow plasma channel.  We
assume $n_c/n_w \ll 1$, and $n_c \ln \left( {r_c}/{ r_{\rm
min}}\right) < n_w \ln \left( 1 + \lambda_D/r_c\right) \approx n_w
\beta_{\rm th}/(k_w r_c)$, where the scattering is dominated by
interaction of the beam with the ions in the wall of the plasma
channel.  In this case the scattering rate is
\begin{equation}
   \frac{d\epsilon_n}{d\gamma} = 
   \frac{ r_e Z_w \beta_{\rm th}}
{2(E_z/E_w) r_c k_\beta\gamma } 
 ,
\end{equation}
where $E_z$ is given by Eq.~\eqref{eq:ez}.  With a focusing force that
maintains constant beam density, $\sigma_x^2 = \epsilon_n/(\gamma
k_\beta) = \textrm{constant}$, the emittance growth is
\begin{equation}
\epsilon_{nf} = \left[ \epsilon_{ni}^2  +   
\frac{ \sigma_x^2 r_e Z_w \beta_{\rm th}}
   { (E_z/E_w) r_c} \left( \gamma_f - \gamma_i \right)
\right]^{1/2} .
\label{eq:emit-const-nb-hollow}
\end{equation}
For typical parameters, $E_z/E_w \sim a^2 \lesssim 1$,
$k_w\sigma_x \sim 1$, $k_wr_c \sim 1$, $Z_w \sim 1$, and $\beta_{\rm
th} \sim 10^{-3}$.  Hence, for $\epsilon_{nf} \gg \epsilon_{ni}$ and
$\gamma_f \gg \gamma_i$, $\epsilon_{nf} \sim (\gamma_f r_e \beta_{\rm
th}/k_w)^{1/2}$.  For high-energy physics applications $\gamma_f =
10^6$, and if the laser-plasma accelerator is operating at a density
of $n_w = 10^{17}~{\rm cm}^{-3}$, then $\epsilon_{nf} \sim 10^{-8}$~m.
This emittance growth is acceptable for high-energy physics
applications.  By contrast, Eq.~\eqref{eq:emit-const-n_b} implies
$\epsilon_{nf} \sim 10^{-6}$~m for similar parameters.

\subsection{Constant focusing force in near-hollow plasma channel}

For a near-hollow plasma channel, the beam density need not be constant and a
constant focusing force can be considered.  In this case, the beam
pinches as it accelerates, however, beam-induced plasma blow-out is
not an issue since the beam is propagating in an ion channel and
on-axis electrons are not required to provide focusing.  Furthermore,
the beam induced wake (beam loading) excited in the channel walls will
be determined by the total beam charge, and not the peak beam density.
The on-axis peak density can increase until limits imposed by ion
motion.

For constant focusing, the rate of scattering is
\begin{equation}
   \frac{d\epsilon_n}{d\gamma} = 
   \frac{ r_e Z_w \beta_{\rm th}}
{2(E_z/E_w) r_c \kappa \gamma^{1/2} } 
 ,
\end{equation}
and the emittance growth is
\begin{equation}
   \Delta \epsilon_n = 
   \frac{ r_e Z_w 
\beta_{\rm  th}}{(E_z/E_w) r_c \kappa}
\left( \gamma_f^{1/2} - \gamma_i^{1/2} \right) 
.
\end{equation}
For typical laser-plasma parameters, $E_z/E_w \sim a^2 \lesssim 1$,
$k_wr_c \sim 1$, $Z_w \sim 1$, and $\beta_{\rm th} \sim 10^{-3}$.
Hence, for $\gamma_f \gg \gamma_i$, $\Delta \epsilon_{n} \sim r_e
\beta_{\rm th} \gamma_f^{1/2} k_w/\kappa$.  For high-energy physics
applications $\gamma_f = 10^6$, and if $\kappa/k_w = 10^{-3}$ (to
prevent ion motion), then $\Delta \epsilon_{n} \sim 10^{-12}$~m,
which is orders of magnitude smaller than that given by
Eqs.~\eqref{eq:emit-const-n_b} or \eqref{eq:emit-const-nb-hollow}.

\section{Summary and Conclusions}

In this work, we have described a method to independently control the
focusing and accelerating forces provided by a plasma accelerator by
using a near-hollow ($n_c/n_w\ll 1$) plasma channel.  The accelerating
wakefield is determined by the wall density $n_w$.  In the limit
$n_c/n_w > a_0^2(r_c)/(8 \gamma_p^2)$ the channel density $n_c$
determines the focusing wakefield, and $k_\beta = k_c/\sqrt{2\gamma}$.
This control of the focusing field allows matched propagation of a
witness electron beam accelerated by the wakefield driven by an
intense laser pulse or relativistic beam.
In principle, this channel geometry can provide perfectly matched
propagation of a high-energy beam, reducing emittance growth.
Emittance growth via Coulomb scattering is mitigated using this
transverse plasma profile, enabling high-energy physics applications.
Use of a near-hollow plasma channel removes the need for low density
bunches $n_b<n_0$ for control of the focusing forces, thereby allowing
strong focusing to be applied (in the quasi-linear regime of
laser-plasma acceleration), further reducing the emittance growth.

The near-hollow plasma channel in the limit described above, in which the
focusing is provided by the channel ions $n_c$, is applicable to
accelerating and focusing electron beams.  For positron beams,
required for high-energy physics applications, one can consider
operating in the limit of a hollow plasma channel and the focusing is
provided externally (e.g., permanent magnets).  In this regime,
electrons and positrons can be accelerated in a nearly symmetrical
fashion.

In this analysis we have neglected ion motion.  Ion motion in the
channel due to the presence of a beam can estimated as $\Delta r/r_0
\sim Z_c (m_e/M_i) (n_b/n_w) (k_w c \Delta t)^2$, where $r_0$ is the
channel ion position, $\Delta r$ is the displacement of the ion during
interaction with the witness beam, and $\Delta t$ is the interaction
time between the witness beam and a background ion.  For resonant
beams $k_w c \Delta t \lesssim 1$.  Sufficiently weak focusing should
be provided such that the beam density satisfies $Z_c
(m_e/M_i)(n_b/n_w) \ll 1$, and the motion of ions in the channel is
negligible.

 Other non-ideal effects may influence the wakefield structure.  For
 example, finite wall thickness in the channel (i.e., a finite
 gradient in plasma density at the wall) will result in a slow decay
 of the wakefield, due to mode coupling at $k_w (r) = k$, with
 characteristic length scale $L_d \sim k_w/(\partial_{r}k_w)$,
 \cite{Shvets96,*Shvets99} typically orders of magnitude larger than
 $k_w^{-1}$.  This analysis assumed a linear plasma response, valid
 for $\vert E_z \vert < E_w$.  For larger laser intensities at the
 channel wall such that $\vert E_z \vert \gtrsim E_w$, wall motion and
 other nonlinear effects will contribute to the wakefields.  Driver
 mis-alignment will also result in excitation of additional mode
 structure.  \cite{Schroeder99a} These modes could introduce
 transverse instabilities (hosing and beam break-up), however, since
 the drivers are short comparted to the plasma wavelength, such
 instabilities will be suppressed.

Although the example presented in Fig.~\ref{fig:wake} considered a
laser driver, a beam driver will produce a similar wakefield structure
in a near-hollow plasma channel with the wakefield phase velocity
approximately the drive beam velocity.\cite{Schroeder99a} Use of
relativistic beams would enable the focusing to be dominated by the
channel ions for lower channel densities.

\begin{acknowledgments}
This work was supported by the
Director, Office of Science, Office of High Energy Physics, of the
U.S.~Department of Energy under Contract No.\  DE-AC02-05CH11231.
\end{acknowledgments}


%

\end{document}